\documentclass{aastex631}
\pdfoutput=1 
\usepackage{amsmath,amstext,amssymb}
\usepackage[figure,figure*]{hypcap}
\usepackage{graphicx} 
\usepackage{multirow}
\usepackage{natbib}
\usepackage{hyperref}
\usepackage[utf8]{inputenc}
\usepackage{xspace}
\usepackage{cancel}

\newcommand{\form}{H$_2$CO\xspace}
\newcommand{\twco}{$^{12}$CO\xspace}
\newcommand{\thco}{$^{13}$CO\xspace}
\newcommand{\lone}{3$_{03}$-2$_{02}$\xspace}
\newcommand{\ltwo}{3$_{21}$-2$_{20}$\xspace}
\newcommand{\lthr}{3$_{22}$-2$_{21}$\xspace}
\newcommand{\red}[1]{}
\newcommand{\twelveco}{$^{12}$CO$\;$2-1\xspace}
\newcommand{\dor}{30$\;$Doradus\xspace}
\newcommand{\kms}{km$\;$s$^{-1}$\xspace}
\newcommand{\uva}{\affiliation{Astronomy Department, University of Virginia, 530 McCormick Rd, Charlottesville, VA 22904, USA}}

\begin{document}

\title{A radial decrease in kinetic temperature measured with \form in \dor}

\author[0000-0002-4663-6827]{R\'emy Indebetouw}
\uva
\affiliation{National Radio Astronomy Observatory, 520 Edgemont Road, Charlottesville, VA 22903, USA}

\author[0000-0002-7759-0585]{Tony Wong}
\affiliation{Department of Astronomy, University of Illinois, Urbana, IL 61801, USA}

\author[0000-0003-3229-2899]{Suzanne Madden}
\affiliation{Departement d'Astrophysique AIM/CEA Saclay, Orme des Merisiers, F-91191 Gif-sur-Yvette, Franc}

\author[0000-0003-2248-6032]{Marta Sewi{\l}o}
\affiliation{Exoplanets and Stellar Astrophysics Laboratory, NASA Goddard Space Flight Center, Greenbelt, MD 20771, USA}
\affiliation{Department of Astronomy, University of Maryland, College Park, MD 20742, USA}
\affiliation{Center for Research and Exploration in Space Science and Technology, NASA Goddard Space Flight Center, Greenbelt, MD 20771}

\author[0000-0001-6326-7069]{Julia Roman-Duval}
\affiliation{Space Telescope Science Institute, 3700 San Martin Drive, Baltimore, MD, 21218, USA}

\author[0000-0002-5635-5180]{Melanie Chevance}
\affiliation{Institut f\"ur Theoretische Astrophysik, Zentrum f\" ur Astronomie der Universit\"at Heidelberg, Albert-\"Uberle-Str.~2, 69120 Heidelberg, Germany}
\affiliation{Cosmic Origins Of Life (COOL) Research DAO, \href{https://coolresearch.io}{https://coolresearch.io}}

\author{Monica Rubio}
\affiliation{Departamento de Astronomía, Universidad de Chile, Casilla 36-D, Santiago, Chile}

\begin{abstract}
Feedback from star formation is a critical component of the evolution of galaxies and their interstellar medium.  At parsec  scales internal to molecular clouds, however, the observed signatures of that feedback on the physical properties of CO-emitting gas have often been weak or inconclusive.  We present sub-parsec observations of \form in the \dor region, which contains the massive star cluster R136 that is clearly exerting feedback on its neighboring gas.  \form provides a direct measure of gas kinetic temperature, and we find a trend of decreasing temperature with projected distance from R136 that may be indicative of gas heating by the stars.  While it has been suggested that mechanical heating affects \form-measured temperature, we do not observe any correlation between T$_{K}$ and line width.  The lack of an enhancement in mechanical feedback close to R136 is consistent with the absence of a radial trend in gravitational boundedness seen the ALMA CO observations. 
Estimates of cosmic ray flux in the region are quite uncertain but can plausibly explain the observed temperatures, if R136 itself is the dominant local source of energetic protons.  The observations presented here are also consistent with the \form-emitting gas near R136 being dominated by direct radiation from R136 and photoelectric heating in the photodissociation regions.
\end{abstract}

\section{Introduction}

Stellar feedback is an essential component of molecular cloud evolution and star formation.  Self-regulated star formation models successfully describe the state of star formation, dense, and diffuse interstellar gas on kiloparsec and galaxy-wide scales, especially on timescales longer than an intermediate to massive stellar lifetime $\gtrsim$10$^7$yr \citep[e.g.][]{OML,KLM}.  However, uncertainty remains about which types of feedback (mechanical, thermal, radiative) act on size scales smaller than a giant molecular cloud (GMC), and timescales less than a few million years \citep{chevance23}.  The \dor region of the Large Magellanic Cloud (LMC) contains one of the nearest massive compact stellar clusters (R136), whose many OB and Wolf-Rayet stars produce prodigious radiation and winds.  Assessments of which type of feedback is most effective on different size and time scales in \dor have come to somewhat different conclusions depending on the details of how that feedback is quantified \citep{pellegrini11,lopez11}.  

In recent years, molecular gas has been observed by ALMA with subparsec resolution in \dor and other star formation regions in the Magellanic Clouds, and the properties of those gas clumps have been analysed for signatures of feedback.  There is some {\em morphological} evidence of very localized feedback associated with individual massive young stellar objects (YSOs).  In N159 for example, individual outflow cavities are observed, and small wind- or HII-region created holes, and gas on the edges of the holes has high brightness temperature CO emission \citep{fukui15,saigo17}. 

However, at least with the modest sample size published to date, there is not strong evidence in the gravo-kinetic properties of parsec-sized clumps that would implicate feedback rather than merely higher surface density.
For example, in \dor, \citet{wong22} find that clumps defined by $^{12}$CO and $^{13}$CO 2-1 emission in \dor have higher velocity dispersions closer to the central R136 cluster, but also have higher surface densities, and are thus no less bound than clumps further away; those observations show no clear signs that feedback from R136 has changed the balance of kinetic and gravitational energy in the molecular gas within 50pc.  Similar CO data in the less evolved N159 massive star formation region south of \dor reveals clumps with lower velocity dispersions than in \dor, but again the virial ratio  of kinetic to gravitational energies is not statistically significantly different \citep{nayak18}, nor do the N159W and E regions appear significantly different in CO-emitting clump properties, despite N159E being surrounded by an HII region and arguably more affected by main-sequence stellar feedback.  Even further south, the quiescent molecular ridge contains no massive stellar clusters, and CO-emitting clumps have even lower velocity dispersions.  The clumps in the quiescent ridge do have statistically lower virial ratios than in \dor (lower kinetic to gravitational energy) although the distributions overlap significantly \citep{finn22}.  The gravo-kinetic properties of clumps traced by CO appears to demonstrate at best weak evidence for feedback on parsec scales.  It is possible that a statistically larger cloud sample may be able to more clearly separate the effects of surface density and feedback.

Measurement of the gas temperature offers an alternative diagnostic of feedback, of particular interest in dense clumps which could potentially form new stars.  Elevated temperatures in star-forming clumps can affect the fragmentation of that gas, and potentially the resulting  stellar initial mass function \citep{heating_IMF}.

Formaldehyde (\form) emission is particularly sensitive to clump kinetic temperatures in dense gas:
\form is a slightly asymmetric rotor, so radiative dipole transitions that change the second rotational quantum number K are nearly forbidden, and the relative populations of the different "K-ladders" are determined almost entirely by collisions.  Thus, it is expected that the relative population of the same J-level of different K-ladders, or the relative strength of transitions from the same J and different K, including \ltwo/\lone and \lthr/\lone, should be strongly sensitive to the kinetic temperature, and only weakly to the collider volume density and  \form column density, as long as the collider volume density is high enough to bring the excitation temperature close to the kinetic temperature.
This temperature-sensitivity, and the similarity of the LTE and non-LTE solutions for reasonable density ranges (10$^4\lesssim$n$_(H_2)\lesssim$10$^6$) has been demonstrated by many authors including \citet[][Fig.~13]{mangum93}, \citet[][Fig.~6]{ginsburg16}, \citet[][Fig.~5]{tang21}.  \citet{tang17} used \form to measure a kinetic temperature of 63.3$^{+18}_{-15}$K in a single 30$\arcsec$ pointing in \dor, the highest of their 6 LMC pointings. 

In \S\ref{observations} we describe the \form observations of \dor.
Subsequent sections describe how those data are used to calculate kinetic temperature (\S\ref{Tkin}), how the results compare to similar analyses in other environments (\S\ref{results}), and comparison with CO and implications for feedback (\S\ref{discussion}).

\section{Data Analysis}
\label{observations}
\subsection{ALMA data}

\begin{figure}
    \centering
    \includegraphics{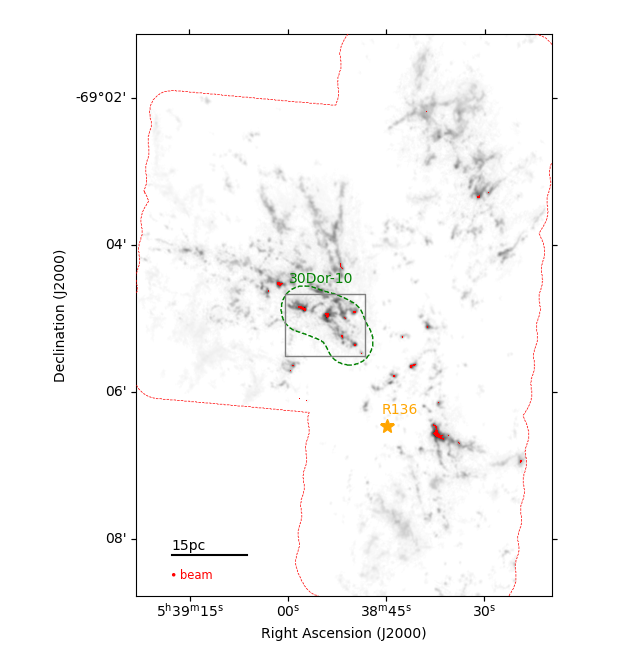}
    \caption{\form\lone integrated intensity ($>$0.8K$\;$km$\;$s$^{-1}\simeq$1$\sigma$) in red overlaid on \twco2-1 integrated intensity in grayscale with a square root stretch. Both lines are integrated from 210-282$\;$km$\;$s$^{-1}$ LSRK. The observed outline is shown in \red{dotted} red, and the location of R136 marked with a yellow star.  The smaller region in cloud 30Dor-10 observed at high resolution is marked in green, and the zoom region of Figure~\ref{fig:http_highres} in gray.}
    \label{fig:emission_on_12co}
\end{figure}

Two datasets are used in this paper as shown in Figure~\ref{fig:emission_on_12co}.  The first, ALMA project 2019.1.00843.S \citet{wong22} covers a wide region at 1.75\arcsec$\;$(0.4pc) resolution.  The second, primarily from ALMA project 2013.1.00346.S \citet{indebetouw20} covers a small central region at 0$\farcs$46$\times$0$\farcs$39 = 0.12$\;$pc$\times$0.098$\;$pc resolution.  Both datasets include the \form transitions listed in Table~\ref{tab:transitions}, as well as 232$\;$GHz continuum.

An approximately 60$\;$pc$\times$90pc region of \dor was observed by ALMA project 2019.1.00843.S between September and December 2019. The data and their calibration are described in detail in \citet{wong22}. The three \form transitions listed in Table~\ref{tab:transitions} were observed in individual spectral windows centered on the lines.
The data were calibrated with the ALMA data processing pipeline Pipeline-CASA56-P1-B \citep{hunter} and Common Astronomy Software Applications (CASA; http://casa.nrao.edu) 5.6.1-8 \citep{casapaper}.  The region was observed in 5 tiles, each calibrated and imaged separately, and then the 5 images were mosaiced together, weighted by their relative sensitivity as a function of position. Only the 12m visibilities were used for the analysis presented here, but peak and integrated clump source densities in the three \form maps were compared to the 7m images, and 95--100\% of the 7m flux is recovered by the higher resolution image, so we are confident in having recovered all detectable emission.
The visibilities were imaged at 1.75\arcsec$\;$(0.4pc)$\;\times\;$0.5$\;$km$\;$s$^{-1}$ resolution, using {tclean} with auto-multithresh auto-masking parameters as used by the ALMA imaging pipeline for 12m data.  The resulting images have sensitivity per channel of $\sigma\simeq$12$\;$mJy$\;$bm$^{-1}$=0.1K. For typical linewidths of $\sim$1$\;$km$\;$s$^{-1}$ the integrated line intensity noise is 0.1$\;$K$\;$km$\;$s$^{-1}$ = 9$\times$10$^{-20}$erg$\;$s$^{-1}\;$cm$^{-2}\;$bm$^{-1}$.

\begin{table}[b]
    \centering
    \caption{\form transitions observed, with their rest frequences and the visibility channel separation in the LSRK frame.}
    \begin{tabular}{c|c|c}
        transition & $\nu_0$ & $\Delta$v \\\hline\hline
         \form 3$_{03}$-2$_{02}$ & 218.22219 GHz & 0.336\kms \\
         \form 3$_{22}$-2$_{21}$ & 218.47532 GHz & 0.335\kms \\
         \form 3$_{21}$-2$_{20}$ & 218.76007 GHz & 0.335\kms \\
    \end{tabular}
    \label{tab:transitions}
\end{table}

\begin{figure}
    \centering
    \includegraphics{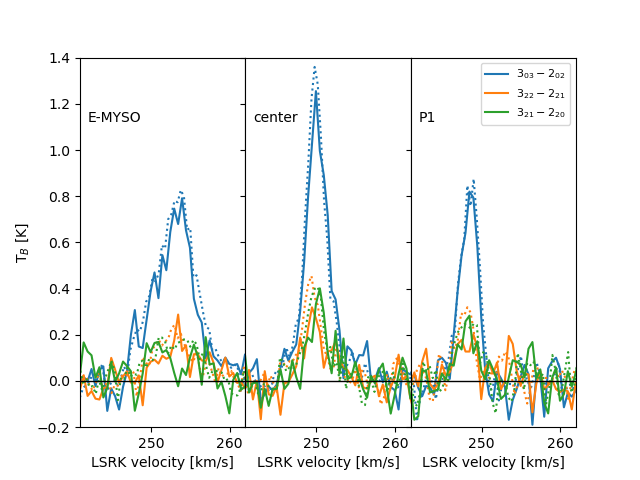}
    \caption{Spectra of the three clumps discussed in more detail in \S\ref{sec:zoom} and Figure~\ref{fig:http_ratio_zooms}. Solid lines are from the 1$\farcs$75 resolution data and dotted from the 0$\farcs$6 resolution data.
    \label{fig:spectra}}
\end{figure}

Figure~\ref{fig:emission_on_12co} shows the distribution of \form and \twco2-1 emission. Figure ~\ref{fig:spectra} shows \form spectra of three clumps in the 30Dor-10 cloud spanning a range from closer to R136 and highly irradiated, to further and more shielded. The same three clumps are highlighted in \S\ref{sec:zoom}.
Integrated intensities were calculated using a threshold mask I(\lone)$>$30mJy$\;$bm$^{-1}$ = 0.25K $\simeq$ 2.5$\sigma$, binary-dilated with a 3-pixel radius (1 pixel = 0$\farcs$5). \form is associated only with the brightest \twco clumps; the association between the species is analyzed in detail below. 

The center of the 30Dor-10 cloud \citep{johansson}, $\sim$20pc north-east of R136, was observed at much higher resolution by project 2013.1.00346.S between June and Sept 2015 (The observed region is marked in Fig.~\ref{fig:emission_on_12co} in green).  The details of those data and their calibration are described in \citet{indebetouw20}.  Visibilities were calibrated with the Cycle3R1 ALMA pipeline included in CASA 4.3.1.  Calibrated visibilities from 2013.1.00346.S, 2019.1.00843.S, and 2011.0.00471.S (an older project described in \citet{indebetouw13}, which covers a somewhat larger part of the 30Dor-10 cloud at lower angular resolution of 1$\farcs$9.) were imaged in CASA 6.6.1 with Briggs weighting robust=2 (nearly the same as natural weighting) and a Gaussian uv taper of 
0.37$\arcsec$, the smoothed (slightly) to a 0$\farcs$6 = 0.15$\;$pc beam and rms noise of 2.5$\;$mJy$\;$bm$^{-1}$ = 0.18K per 0.4$\;$km$\;$s$^{-1}$ channel.
For typical linewidths of $\sim$1$\;$km$\;$s$^{-1}$ the integrated line intensity noise is 0.3$\;$K$\;$km$\;$s$^{-1}$.
Integrated and convolved peak fluxes of clumps are 95--100\% of the lower resolution data from 2019.1.00843.S, indicating that resolved out flux is again negligible, and the emitting structures are smaller than the $\sim$2\arcsec=0.5pc largest recoverable scale of the higher resolution data.  

\subsection{Clump Segmentation}

For analysis, the \form\lone emission is segmented into clumps using {\tt astrodendro}, min\_value = 3$\sigma$ = 0.3K, min\_delta = 2.5$\sigma$ = 0.25K, min\_npix = 1$\;$beam area = 14 square pixels.   The min\_npix parameter is the smallest allowed clump, and the min\_delta parameter is the difference between a peak and the valley between two peaks that determines when those peaks are considered separate objects. The parameters matter very little for subsequent analysis since only two of the clumps break into substructure, and only into 3 sub-clumps each. 
The clump boundaries in position-position-velocity space determined from \lone are then applied to the other two fainter lines to sum their emission.
Fifty-three structures are found in \lone, and 29 have signal-to-noise greater than 3 in the fainter \ltwo or \lthr line.  The clumps and their properties in \form and continuum are listed in Table~\ref{table:clumps}.

Continuum emission at 232GHz in star-forming regions at these scales is a combination of free-free emission from ionized gas, and warm dust.  Compact emission from both mechanisms is often associated with embedded young stellar objects, and thus useful to qualitatively assess the embedded star formation content of \form clumps.
Continuum at 232GHz from project 2019.1.00843.S was imaged in CASA 6.5.4.9, with multi-scale deconvolution at scales of 0,5,11, and 21$\times$ the 0$\farcs$5 pixel size, and then convolved to the same 1$\farcs$75 resolution as the \form cubes.
The three central tiles "b", "c", and "d" were deconvolved together, and outlying tiles "a" and "e" were deconvolved separately and then linearly mosaiced with the central image.  The reason for this is because the central tiles contain significant diffuse emission which is resolved out by the interferometer. We used total power SIMBA 1.2mm observations \citep{ferreira} to recover that emission, both by using the total power as an initial model in the deconvolution, and by feathering the deconvolved interferometric image with the total power image.  The recovery of diffuse emission makes no difference to any conclusions presented here, because the continuum directly associated with compact \form clouds is also very compact.

\begin{table}
\begin{rotatetable}
\setlength{\tabcolsep}{0.25em}
\movetableright=-11.5cm
\caption{Properties of \form-emitting clumps. $\sigma$ are the second moments of \form $3_{03}$-$2_{02}$ in each direction, F are line-integrated fluxes, and I are peak intensities.  For the \form lines, $1,2,3$ refer to $3_{03}$-$2_{02}$, $3_{22}$-$2_{21}$, and $3_{21}$-$2_{20}$, respectively.  $^{13}$CO is the 2-1 transition, "ci" refers to 232GHz continuum, interferometer only, and "c" to the combined ALMA interferometric and SIMBA total power data.  dF is the 1$\sigma$ noise level in the line cubes, multiplied by the square root of the number of measurements in the clump, i.e. the number of beams in its angular extent times number of velocity channels. 
\label{table:clumps}}
\vspace*{-3ex}
\begin{tabular}{cccc|ccc|cccc|cccccccc|ccc|ccc}
\\\hline
\renewcommand{\colhead}{}
\colhead{i} &
\colhead{R.A.} &
\colhead{Decl.} &
\colhead{v$_{\rm LSRK}$} & 
\colhead{$\sigma_{\rm ra}$} &
\colhead{$\sigma_{\rm dec}$} &
\colhead{$\sigma_{\rm v}$} &
\colhead{$F_{\rm 1}$} &
\colhead{$F_{\rm 2}$} &
\colhead{$F_{\rm 3}$} &
\colhead{dF} &
\colhead{$I_{\rm 1}$} &
\colhead{$I_{\rm 2}$} &
\colhead{$I_{\rm 3}$} &
\colhead{I$_\mathrm{13}$} & 
\colhead{dI$_\mathrm{13}$} & 
\colhead{I$_{\rm ci}$} & 
\colhead{dI$_{\rm ci}$} & 
\colhead{I$_c$} &
\colhead{T$_{\rm K,fit}$} & 
\colhead{dT$_K^+$} & 
\colhead{dT$_K^-$} &
\colhead{N$_{\rm fit}$} & 
\colhead{dN$^{\rm +}$} & 
\colhead{dN$^{\rm -}$} \\
\colhead{} & 
\colhead{(deg)} & 
\colhead{(deg)} & 
\colhead{($\mathrm{km\,s^{-1}}$)} & 
\colhead{($\mathrm{pc}$)} & 
\colhead{($\mathrm{pc}$)} & 
\multicolumn{1}{c|}{($\mathrm{km\,s^{-1}}$)} &
\multicolumn{4}{c|}{($\mathrm{K\,km\,s^{-1}\,pc^{-2}}$)} &
\multicolumn{8}{c|}{($\mathrm{K}$)} &
\multicolumn{3}{c|}{($\mathrm{K}$)} &
\multicolumn{3}{c}{($\mathrm{cm^{-2}}$)}
\\\hline
1 & 84.60961 & -69.05177 & 252.22 & 0.312 & 0.261 & 0.61 & 1.04 & 0.18 & 0.14 & 1.37 & 0.438 & 0.350 & 0.032 & 8.02 & 1.53 & 0.023 & 0.008 & 0.023 & 32.5 & 12.5 & 10.0 & 13.4 & 3.3 & 0.4 \\
2 & 84.62148 & -69.10761 & 250.22 & 0.451 & 0.221 & 0.34 & 0.36 & 0.05 & 0.04 & 0.75 & 0.306 & 0.382 & 0.021 & 6.45 & 1.74 & 0.018 & 0.006 & 0.029 & 0.0 & 0.0 & 0.0 & \nodata & \nodata & \nodata \\
3 & 84.62629 & -69.06109 & 252.22 & 0.281 & 0.224 & 0.25 & 0.28 & 0.04 & 0.07 & 0.76 & 0.283 & 0.278 & 0.018 & 7.26 & 1.49 & 0.012 & 0.005 & 0.015 & 45.0 & 30.0 & 17.5 & 13.1 & 0.4 & 0.3 \\
4 & 84.62850 & -69.10595 & 250.22 & 0.206 & 0.176 & 0.22 & 0.21 & 0.06 & 0.03 & 0.85 & 0.492 & 0.338 & 0.014 & 5.42 & 0.71 & 0.015 & 0.004 & 0.030 & 50.0 & 30.0 & 17.5 & 13.2 & 0.4 & 0.3 \\
5 & 84.63024 & -69.03720 & 249.72 & 0.349 & 0.138 & 0.30 & 0.23 & 0.04 & 0.04 & 0.84 & 0.318 & 0.266 & 0.016 & 4.46 & 1.07 & 0.061 & 0.005 & 0.061 & 37.5 & 20.0 & 25.0 & 13.2 & 0.3 & 0.4 \\
6 & 84.63412 & -69.03803 & 249.72 & 0.194 & 0.337 & 0.44 & 0.32 & 0.14 & 0.09 & 0.80 & 0.605 & 0.487 & 0.020 & 6.33 & 1.22 & 0.041 & 0.007 & 0.041 & 100.0 & 100.0 & 40.0 & 13.3 & 0.6 & 0.3 \\
7 & 84.63468 & -69.12303 & 249.22 & 0.170 & 0.108 & 0.21 & 0.06 & 0.02 & 0.02 & 0.72 & 0.246 & 0.250 & 0.009 & 3.49 & 0.57 & 0.020 & 0.003 & 0.024 & 60.0 & 50.0 & 22.5 & 13.1 & 0.4 & 0.3 \\
8 & 84.63514 & -69.09845 & 248.22 & 0.266 & 0.169 & 0.25 & 0.27 & 0.05 & 0.03 & 0.86 & 0.375 & 0.295 & 0.017 & 6.28 & 1.29 & 0.032 & 0.004 & 0.049 & 0.0 & 0.0 & 0.0 & \nodata & \nodata & \nodata \\
9 & 84.63629 & -69.10553 & 246.22 & 0.786 & 0.747 & 1.39 & 12.39 & 2.35 & 2.36 & 2.14 & 0.631 & 0.650 & 0.094 & 12.22 & 2.53 & 0.100 & 0.018 & 0.114 & 40.0 & 10.0 & 10.0 & 13.4 & 1.8 & 0.3 \\
10 & 84.63629 & -69.10553 & 246.22 & 0.624 & 0.411 & 1.02 & 7.50 & 1.46 & 1.46 & 2.14 & 0.631 & 0.650 & 0.061 & 12.22 & 1.85 & 0.100 & 0.012 & 0.114 & 40.0 & 12.5 & 10.0 & 13.4 & 1.8 & 0.3 \\
11 & 84.63629 & -69.10553 & 246.22 & 0.684 & 0.730 & 1.16 & 10.75 & 2.02 & 2.09 & 2.14 & 0.631 & 0.650 & 0.081 & 12.22 & 2.40 & 0.100 & 0.015 & 0.114 & 40.0 & 10.0 & 10.0 & 13.4 & 1.8 & 0.3 \\
12 & 84.63863 & -69.10345 & 245.72 & 0.292 & 0.216 & 0.60 & 1.01 & 0.16 & 0.22 & 1.69 & 0.384 & 0.523 & 0.024 & 5.93 & 0.74 & 0.029 & 0.006 & 0.044 & 40.0 & 10.0 & 12.5 & 13.4 & 1.8 & 0.4 \\
13 & 84.64179 & -69.08137 & 253.22 & 0.480 & 0.200 & 0.36 & 0.50 & 0.09 & 0.04 & 0.75 & 0.247 & 0.294 & 0.025 & 7.41 & 1.50 & 0.032 & 0.006 & 0.042 & 30.0 & 22.5 & 20.0 & 13.1 & 0.4 & 0.3 \\
14 & 84.64640 & -69.10901 & 243.72 & 0.255 & 0.158 & 0.32 & 0.35 & 0.09 & 0.09 & 1.34 & 0.344 & 0.488 & 0.018 & 4.58 & 0.96 & 0.038 & 0.005 & 0.051 & 60.0 & 25.0 & 17.5 & 13.5 & 0.2 & 0.4 \\
15 & 84.65228 & -69.09012 & 249.22 & 0.593 & 0.256 & 0.76 & 2.62 & 0.63 & 0.61 & 1.89 & 0.516 & 0.559 & 0.045 & 11.38 & 2.09 & 0.042 & 0.009 & 0.060 & 50.0 & 17.5 & 12.5 & 13.5 & 3.3 & 0.3 \\
16 & 84.65735 & -69.08346 & 262.22 & 0.292 & 0.149 & 0.36 & 0.38 & 0.08 & 0.10 & 1.10 & 0.282 & 0.419 & 0.019 & 5.83 & 1.01 & 0.014 & 0.005 & 0.023 & 50.0 & 25.0 & 15.0 & 13.4 & 0.3 & 0.4 \\
17 & 84.65848 & -69.10971 & 241.72 & 0.281 & 0.146 & 0.24 & 0.18 & 0.04 & 0.07 & 0.88 & 0.349 & 0.383 & 0.015 & 3.62 & 0.77 & 0.027 & 0.004 & 0.035 & 77.5 & 55.0 & 27.5 & 13.3 & 0.4 & 0.3 \\
18 & 84.66317 & -69.09234 & 244.22 & 0.333 & 0.173 & 0.47 & 0.98 & 0.26 & 0.27 & 1.77 & 0.695 & 0.543 & 0.028 & 7.36 & 1.49 & 0.027 & 0.007 & 0.042 & 60.0 & 22.5 & 15.0 & 13.5 & 0.6 & 0.3 \\
19 & 84.68107 & -69.09915 & 235.72 & 0.331 & 0.150 & 0.24 & 0.19 & 0.03 & 0.03 & 0.68 & 0.280 & 0.282 & 0.016 & 4.95 & 1.08 & 0.016 & 0.004 & 0.021 & 35.0 & 25.0 & 25.0 & 13.1 & 0.4 & 0.4 \\
20 & 84.68380 & -69.08707 & 250.22 & 0.256 & 0.132 & 0.25 & 0.21 & 0.07 & 0.07 & 0.86 & 0.338 & 0.420 & 0.015 & 3.58 & 0.75 & 0.028 & 0.004 & 0.039 & 87.5 & 67.5 & 32.5 & 13.3 & 0.4 & 0.3 \\
21 & 84.68731 & -69.07790 & 250.72 & 0.468 & 0.259 & 0.56 & 1.11 & 0.18 & 0.16 & 1.26 & 0.375 & 0.372 & 0.034 & 6.97 & 1.19 & 0.027 & 0.008 & 0.049 & 0.0 & 0.0 & 0.0 & \nodata & \nodata & \nodata \\
22 & 84.68808 & -69.08526 & 248.72 & 0.368 & 0.150 & 0.39 & 0.62 & 0.13 & 0.17 & 0.99 & 0.295 & 0.458 & 0.025 & 9.38 & 1.71 & 0.084 & 0.005 & 0.098 & 55.0 & 30.0 & 17.5 & 13.3 & 0.3 & 0.4 \\
23 & 84.69431 & -69.07915 & 255.22 & 0.402 & 0.147 & 0.36 & 0.71 & 0.13 & 0.13 & 1.24 & 0.360 & 0.337 & 0.025 & 8.95 & 1.76 & 0.052 & 0.005 & 0.078 & 40.0 & 15.0 & 12.5 & 13.4 & 3.3 & 0.4 \\
24 & 84.69547 & -69.08471 & 248.22 & 0.278 & 0.126 & 0.26 & 0.22 & 0.07 & 0.06 & 0.80 & 0.425 & 0.384 & 0.016 & 4.50 & 0.77 & 0.038 & 0.004 & 0.054 & 0.0 & 0.0 & 0.0 & \nodata & \nodata & \nodata \\
25 & 84.69586 & -69.08332 & 250.22 & 0.341 & 0.246 & 0.50 & 1.11 & 0.27 & 0.35 & 1.60 & 0.427 & 0.550 & 0.031 & 10.36 & 1.94 & 0.048 & 0.008 & 0.068 & 65.0 & 25.0 & 17.5 & 13.4 & 0.6 & 0.3 \\
26 & 84.70403 & -69.07999 & 253.22 & 0.679 & 0.179 & 0.45 & 0.34 & 0.04 & 0.08 & 0.66 & 0.345 & 0.373 & 0.022 & 4.61 & 1.09 & 0.051 & 0.006 & 0.073 & 40.0 & 30.0 & 30.0 & 13.1 & 0.4 & 0.3 \\
27 & 84.70520 & -69.07846 & 249.72 & 0.674 & 0.393 & 0.54 & 2.96 & 0.69 & 0.65 & 1.85 & 0.468 & 0.537 & 0.051 & 14.35 & 2.79 & 0.154 & 0.011 & 0.176 & 47.5 & 15.0 & 12.5 & 13.5 & 3.3 & 0.4 \\
28 & 84.71219 & -69.07304 & 247.72 & 0.331 & 0.155 & 0.34 & 0.23 & 0.02 & 0.03 & 0.71 & 0.338 & 0.506 & 0.017 & 5.39 & 1.07 & 0.021 & 0.005 & 0.038 & 10.0 & 37.5 & 0.0 & 13.1 & 0.4 & 0.4 \\
29 & 84.72036 & -69.07721 & 252.72 & 1.279 & 0.315 & 1.02 & 5.75 & 1.00 & 0.75 & 1.86 & 0.846 & 0.504 & 0.068 & 8.28 & 1.56 & 0.105 & 0.012 & 0.121 & 32.5 & 10.0 & 10.0 & 13.4 & 2.2 & 0.4 \\
30 & 84.72700 & -69.09012 & 257.22 & 0.397 & 0.289 & 0.95 & 0.79 & 0.15 & 0.15 & 0.77 & 0.655 & 0.526 & 0.030 & 6.18 & 1.33 & 0.019 & 0.008 & 0.027 & 42.5 & 27.5 & 17.5 & 13.1 & 0.4 & 0.3 \\
31 & 84.72856 & -69.09137 & 259.22 & 0.363 & 0.286 & 0.44 & 0.41 & 0.06 & 0.12 & 0.83 & 0.519 & 0.529 & 0.022 & 4.92 & 1.09 & 0.017 & 0.007 & 0.025 & 47.5 & 30.0 & 17.5 & 13.2 & 0.3 & 0.4 \\
32 & 84.73474 & -69.07151 & 250.22 & 0.714 & 0.243 & 0.68 & 3.66 & 0.65 & 0.53 & 1.78 & 0.563 & 0.601 & 0.052 & 9.03 & 1.95 & 0.040 & 0.010 & 0.050 & 35.0 & 10.0 & 10.0 & 13.4 & 3.9 & 0.4 \\
33 & 84.74330 & -69.07262 & 251.22 & 0.363 & 0.232 & 0.30 & 0.54 & 0.08 & 0.11 & 1.13 & 0.407 & 0.383 & 0.023 & 7.25 & 1.46 & 0.013 & 0.006 & 0.018 & 37.5 & 17.5 & 12.5 & 13.4 & 0.3 & 0.4 \\
\end{tabular}
\end{rotatetable}
\end{table}

\clearpage
\section{Kinetic Temperature Calculation}
\label{Tkin}

\begin{figure}
    \centering
    \resizebox{\textwidth}{!}{\includegraphics{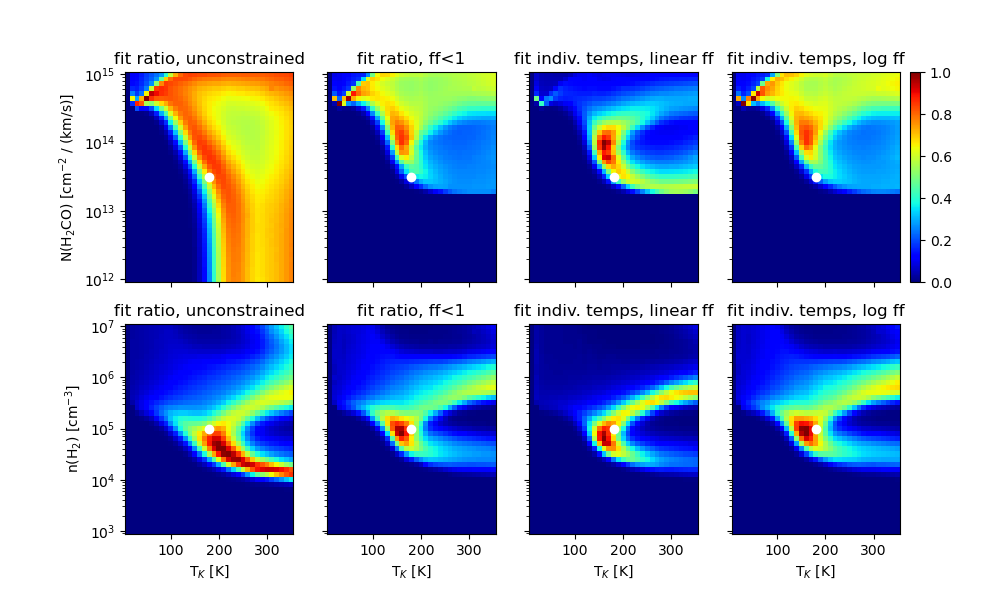}}
    \caption{Probability distributions resulting from fitting three \form lines and uncertainty dT=0.1$\;$K.
    These plots show the result of fitting an arbitrarily chosen set of conditions marked with a white circle, and demonstrates potential systematic offsets in this kind of fitting.
    The offset from the true value is not systematic over parameter space. In columns 1 and 2 two line ratios are fit as a function of H$_2$ density $n$, \form column density $N$, and kinetic temperature $T_K$, and the resulting probability distribution is shown as a function of two variables, marginalized over the third.  
    Although the linear filling factor {\it ff} is not explicitly fit, in column 2 it is constrained to be less than 1 within uncertainty by requiring $T_{model}<T_{measured}-dT$ for all 3 lines.  
    In columns 3 and 4 the three individual intensities are fit along with a filling factor -- in column 3 the {\it ff} prior distribution is uniform in linear space between 0.01 and 1, and in column 4 the prior is uniform in logarithmic space between 10$^{-2}$ and 1. Fitting intensities with a logarithmic {\it ff} prior is the same as fitting line ratios and imposing {\it ff}$<1$.  
    However, there is a systematic bias to high $N$ in the range of allowed fit parameters, which can be reduced by using a linear uniform prior (lower probability of small {\it ff} which correspond to large $N$ and large optical depth $\tau$).
    }
    \label{fig:how2fit}
\end{figure}

We fit the three \form lines using grids of non-LTE models calculated with RADEX \citep{radex}, and tested several different methods to determine $T_K$, as illustrated in Figure~\ref{fig:how2fit}.  For each clump of emission, we calculated $\chi^2$ as a function of $H_2$ volume density $n$, \form column density $N$, kinetic temperature $T_K$, and beam areal filling factor {\it ff}.  Flat priors in log space were used for the physical variables, and either a flat prior in linear or in log space for the filling factor.
Note that the three observed \form lines are all para-\form, so physical conditions can be constrained without assuming an ortho-to-para ratio.

\begin{figure}
    \centering
    \resizebox{\textwidth}{!}{\includegraphics{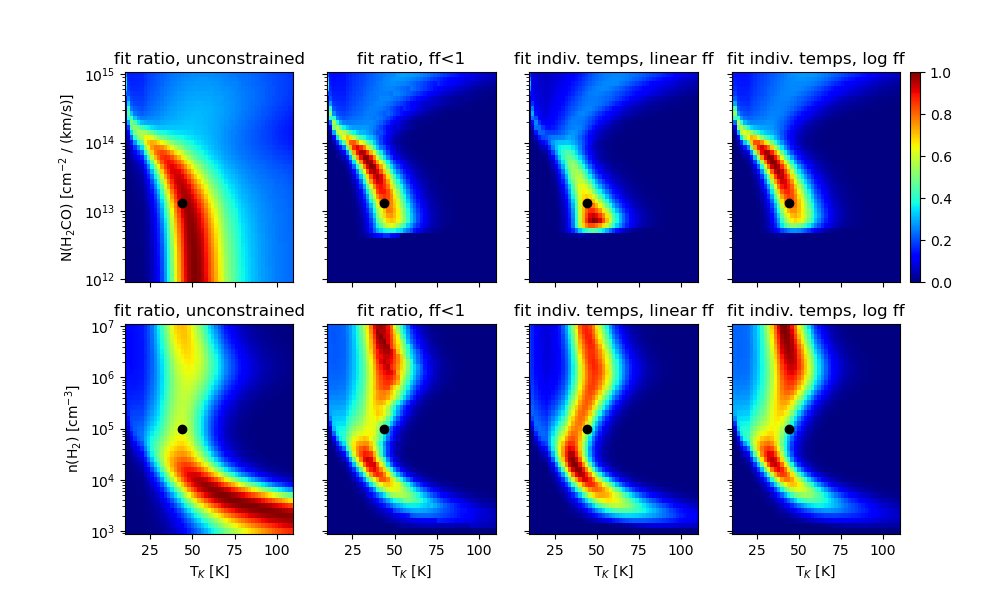}}
    \caption{Probability distributions resulting from fitting the brightest clump in our data. All of the panels and labels are the same as Figure~\ref{fig:how2fit}, but now the black circle marks the LTE derived parameters, to illustrate potential offsets from LTE.  This Figure also highlights the cooler part of parameter space, contrasted with the warmer Figure~\ref{fig:how2fit}.
    }
    \label{fig:how2fitclump}
\end{figure}

We compared the physical conditions corresponding to the peak probability as a function of 4, 3 (filling factor marginalized), 2 (filling factor and $n$ marginalized), and 1 dimension (all but $T_K$ or $N$ marginalized).  All of those methods result in consistent $T_K$ within uncertainties, as expected by the molecular structure considerations that predict predominantly $T_K$ sensitivity. 
Figures~\ref{fig:how2fit} and \ref{fig:how2fitclump} demonstrate the different methods, showing the marginalized probability distributions for each. 
If fitting a filling factor, it is important to restrict the filling factor to be less than 1.  With that restriction, fitting line ratios or the line intensities directly give the same best fit, but the uncertainty ranges differ somewhat depending on the choice of prior for the filling factor:
Fitting the three line brightness temperatures with a log prior on filling factor yields the same posterior probability distributions as fitting the two line ratios and imposing that the measured brightness temperature of \lone is less than that of the model (i.e. excluding a non-physical filling factor $>$1). A linear prior for filling factor results in $\sim$2$\times$ lower expectation values for fitted column density, not surprisingly since a log prior has greater probability of low filling factor, and correspondingly higher fitted column densities, than a linear filling factor prior.  The two methods yield nearly identical expectation values for the kinetic temperature. 

\begin{figure}[h]
    \centering
    \resizebox{3.5in}{!}{\includegraphics{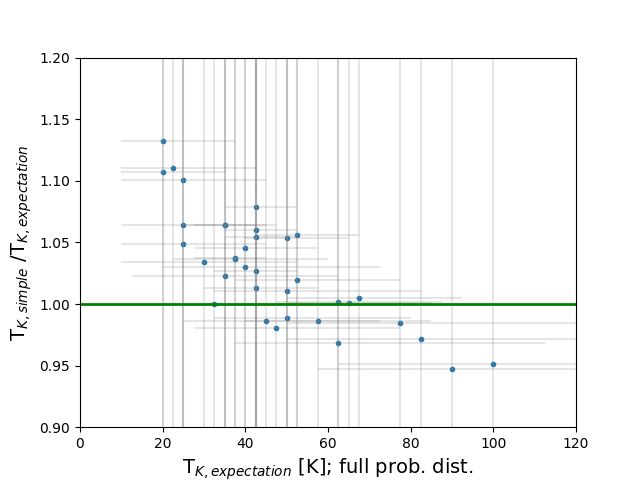}}
    \resizebox{3.5in}{!}{\includegraphics{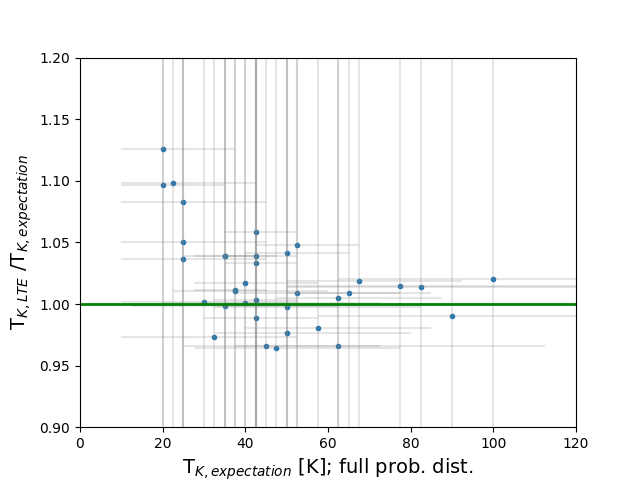}}
    \caption{First panel: Kinetic temperature $T_{K,Tang21}$ following the "simple" conversion function derived from non-LTE models at $n$($H_2$)=2$\times$10$^5$cm$^{-3}$ and N(\form)=10$^{13}\;$cm$^{-2}$, compared to the expectation value of kinetic temperature $T_{K,expectation}$ modeled as a function of a wide range of $n$($H_2$), $N$(\form), $T_K$, and filling factor.  The y axis is the ratio of the "simple" method to the full 4-dimensional method.
    Second panel: the same comparison but with the LTE line ratio.}
    \label{fig:tang_v_fit}
\end{figure}

\citet{tang21} used the average of the \ltwo and \lthr lines, and the ratio of that average to \lone, to calculate $T_K$. 
We perform a similar calculation, using the NLTE averaged-line ratio at fixed N(\form)=10$^{13}\;$cm$^{-2}$ and n($H_2$)=2$\times$10$^5\;$cm$^{-3}$. Those values are  
higher than the N(\form)=10$^{12}\;$cm$^{-2}$ and n($H_2$)=10$^5\;$cm$^{-3}$ used by \citet{tang21} with lower angular resolution data, but are more representative of the full 4-dimensional fitted values that we derive from the 
1$\farcs$75 data.
Figure~\ref{fig:tang_v_fit} shows the comparison between our expectation values for $T_K$ in the 4-parameter model fits, compared to the conversion at fixed N(\form) and N(H$_2$), and compared to the LTE conversion.  There is a modest tendency (20\%) for the "simple" or LTE conversions to yield higher temperatures when $T_K\lesssim$50K, and lower temperatures when $T_K\gtrsim$100K, compared to the full modeling.  For the rest of the analysis of the entire region, we use the full modeling $T_K$ results for clumps (most of which are marginally resolved at 1$\farcs$75 resolution).
For higher-resolution maps of \form temperature distributions we use the \citet{tang21} "simple" method of converting from line ratio to T at fixed N(\form) and n(H$_2$), to avoid nonlinearities in the fitting that can be more pronounced at lower signal-to-noise, and to present temperature trends within clumps that are more transparently derived directly from the line ratios. 

The "simple" conversion of the average \ltwo and \lthr ratio to \lone to $T_K$ can also be less accurate if the optical depth of \lone becomes significant.  To assess this, we calculate the synthesized beam brightness temperature $T_{mb}$=S$_\nu\;\lambda^2$/(2.65$\theta_{arcmin}^2$) where $\lambda$ and $\theta_{arcmin}$ are the wavelength and synthesized beam in arcminutes, respectively. The peak synthesized beam brightness temperature for the brightest clump is 2.1K, much lower than the fitted kinetic temperature of 42K for that clump. All clumps have brightness temperatures $<$7\% of their fitted kinetic temperatures,
so either the optical depth is low, or the filling factor is $\ll$0.1, or the excitation temperature is $\ll$ the kinetic temperature.  Some combination of the three is probably true in this case.

\clearpage
\section{Gas temperature in \dor and other regions}
\label{results}

\begin{figure}
    \centering
    \includegraphics{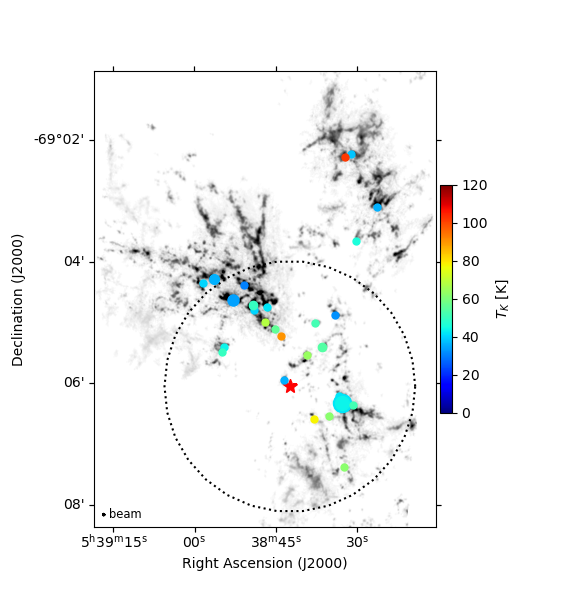}
    \caption{Fitted T$_K$ overlaid on integrated \twelveco as in Fig.~\ref{fig:emission_on_12co}.  The symbol size scales with the clump integrated flux of \lone, and its color scales with temperature. The dotted circle indicates 30pc distance from R136, marked with a red star.  See text for discussion.}
    \label{fig:temp_on_12co}
\end{figure}

\begin{figure}
    \centering
    \includegraphics{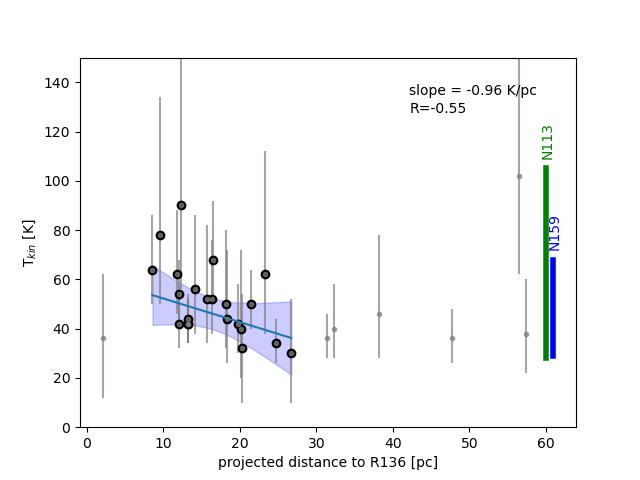}
    \caption{Fitted T$_K$ as a function of projected distance from R136.  The most significant trend is found between 5 and 30pc projected distance; the points in that range are highlighted with a darker symbol, and the best fit to them, and 95\% confidence band are shown here. 
    The point nearest to \dor in projection is seen in silhouette in the optical \citep{kalari}, and likely the true distance significantly exceeds the projected distance. The range of derived temperatures in LMC regions N113 and N159 from \citet{tang21} are shown for comparison.}
    \label{fig:temp_radtrend}
\end{figure}

Fitted kinetic temperature is shown overlaid on \twelveco in Figure~\ref{fig:temp_on_12co}.  There appears to be a trend of decreasing temperature with increasing projected distance from the R136 cluster. Figure~\ref{fig:temp_radtrend} shows the fitted temperatures as a function of projected distance.  The minimum and maximum projected distance were allowed to vary to maximize the significance of the linear fit as quantified by the Pearson's correlation coefficient $R$. That maximization indicates that the most significant trend is within 30pc of R136, indicated on Fig.~\ref{fig:temp_on_12co}, and outside of that radius, a linear model is no better at explaining the variance of the data than a model with no radial dependence. 
The fitted slope depends somewhat on the precise range of points chosen, resulting in an uncertainty of $\pm$0.5$\;$K$\;$pc$^{-1}$.

\citet{tang17} measured $T_K$ from these three \form lines at 6 locations in the LMC with a 30$\arcsec$ beam. Their pointing in \dor is at RA(J2000)=5$^{\rm h}$30$^{\rm m}$49.3$^{\rm s}$, Dec(J2000)=-69$^{\circ}$04$'$44$''$, corresponding to the center of the region observed at higher angular resolution here (Figs~\ref{fig:emission_on_12co} and \ref{fig:http_ratio_zooms}, central panels). They calculate $T_K$=63.3$^{+18}_{-15}$K, the main clump is unresolved in our lower-resolution data, and we calculate $T_K$=50$^{+17}_{-13}$K, in good agreement given the resolution difference. The higher resolution data (Fig~\ref{fig:http_ratio_zooms}) has line ratios corresponding to 50 and 80K in the centers of the two brightest clumps, increasing to $\gtrsim$100K on the clump edges. Some of the other clumps that would have been in the 30$\arcsec$ beam have lower temperatures, so again the measurements are consistent.  
%
The \form temperature ranges also agree well with the other LMC regions N113 and N159 measured at very similar angular resolution to this work, by \citet{tang21} with the same lines.  They found a range of temperatures in N113 of 28-105K (median 49K), and in N159 a range of 29-68K (median 49K).  These ranges agree well with the range found in 30Dor: 20-100K, median 43K, mean 48$\pm$3K (standard error of the mean).  The temperature distributions of the three regions agree within their uncertainties.
%


\begin{figure}
    \centering
    \includegraphics{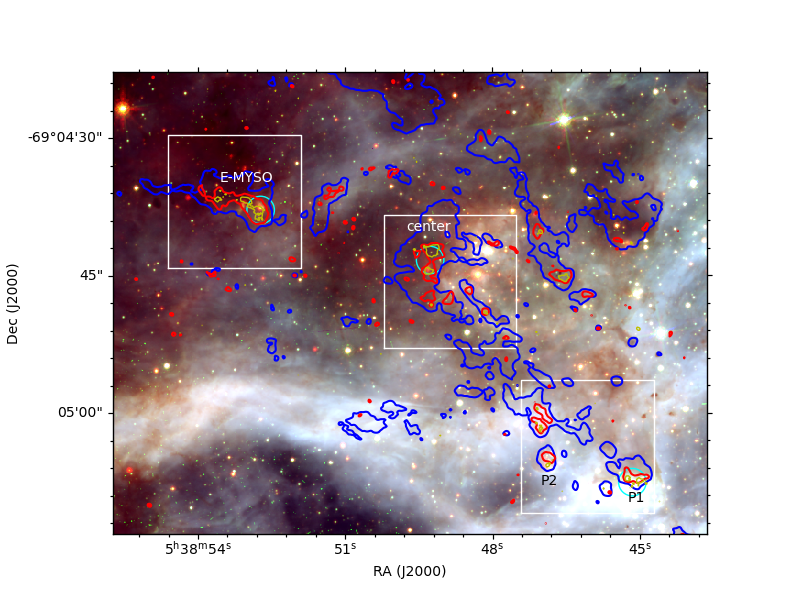}
    \caption{The central region 30Dor-10, 20pc NE of R136. The color image is BGR= equivalent V, R, and H band from the Hubble Tarantula Treasury Project (HTTP). A 2$\sigma$ contour of \thco$\;$2-1 is shown in blue, and the 2$\sigma$ contour of \form \lone is shown in red. Spitzer-identified embedded massive young stellar objects are indicated with cyan circles and the yellow contour indicates 230GHz continuum at 10$^{-4}\;$Jy$\;$bm$^{-1}\;\simeq\;$3$\sigma$.  The \form beam is shown in the lower left in red.  The pillars P1 and P2 are surrounded by gas ionized by R136 to the south. Zooms into three regions and the derived $T_K$ are shown in Fig.~\ref{fig:http_ratio_zooms}.}
    \label{fig:http_highres}
\end{figure}

\begin{figure}
    \centering
    \resizebox{2.9in}{!}{\includegraphics[trim=0 30 0 30, clip]{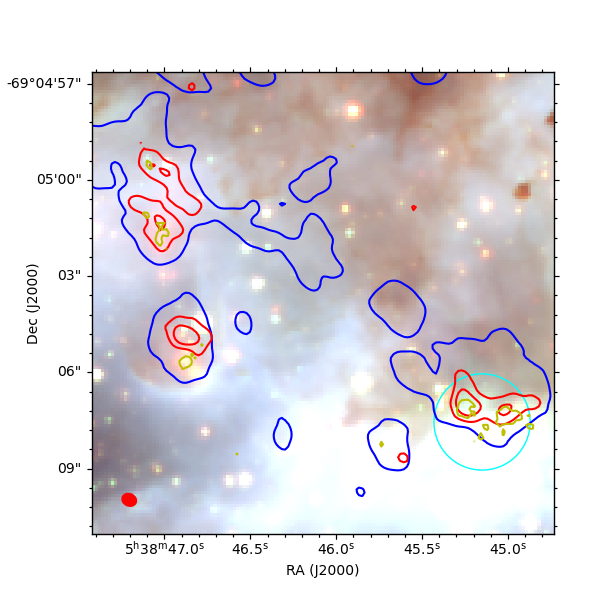}}
    \resizebox{3.2in}{!}{\includegraphics[trim=0 10 0 0, clip]{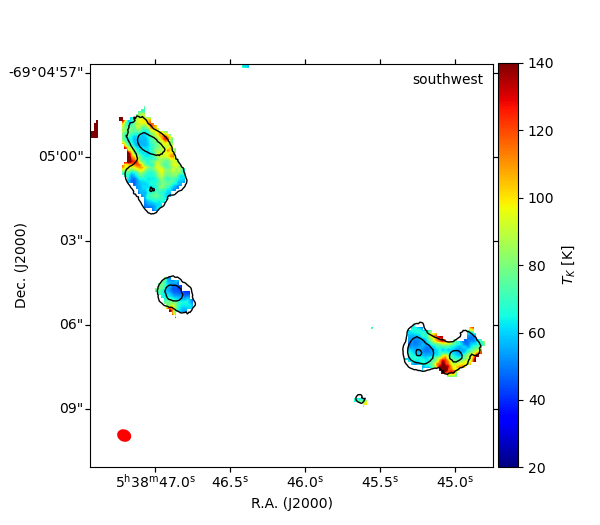}}\\
    \resizebox{2.9in}{!}{\includegraphics[trim=0 30 0 30, clip]{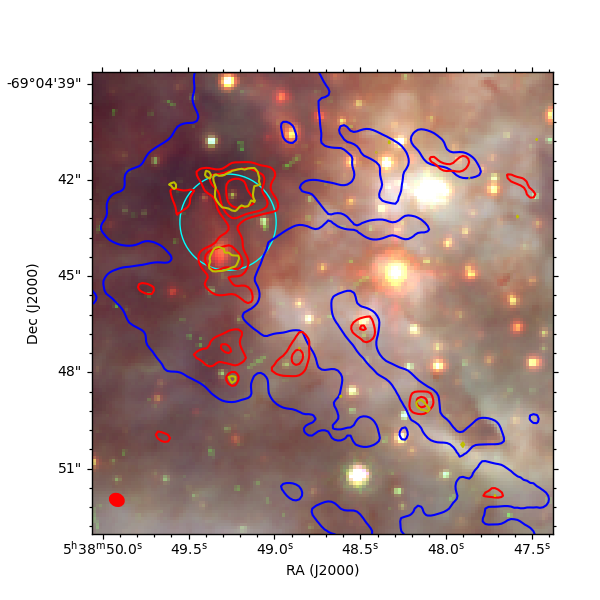}}
    \resizebox{3.2in}{!}{\includegraphics[trim=0 10 0 0, clip]{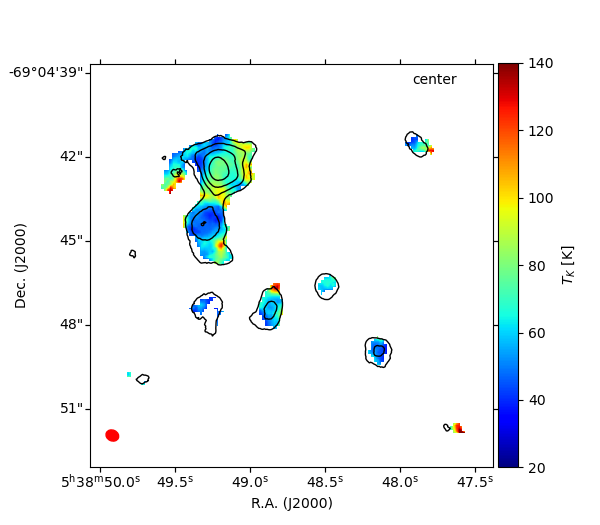}}\\
    \resizebox{2.9in}{!}{\includegraphics[trim=0 30 0 30, clip]{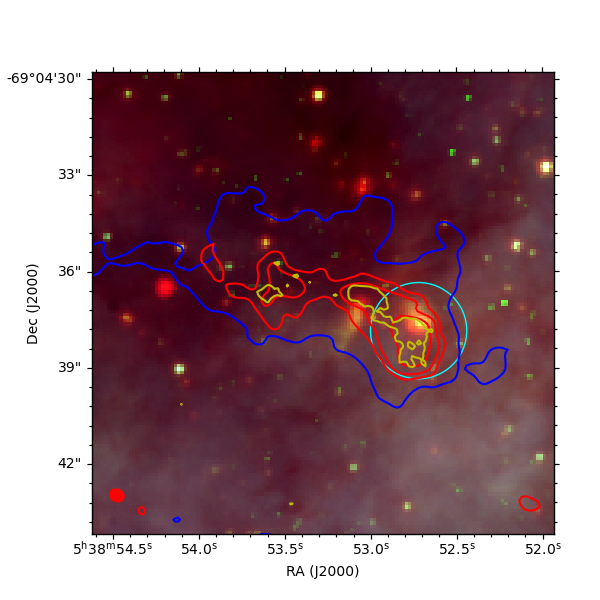}}
    \resizebox{3.2in}{!}{\includegraphics[trim=0 10 0 0, clip]{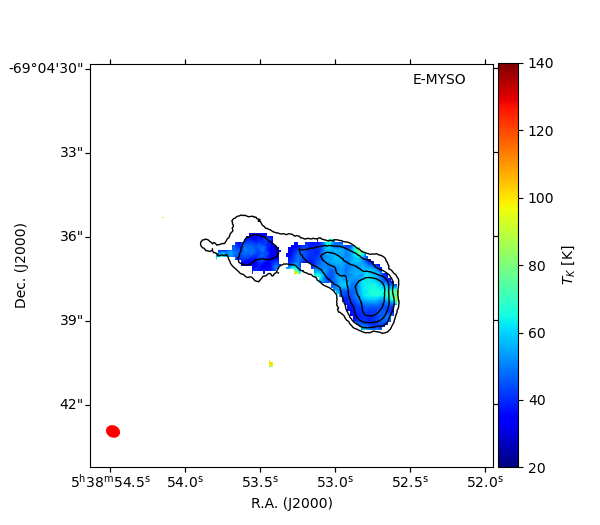}}\\
    \caption{Zooms into the main clumps of 30Dor-10: (top) the pillars on the R136-facing southern edge of the cloud, (middle) the central clumps, and (bottom) the eastern clump "E-MYSO". The left panels show Hubble Tarantula Treasury Project (HTTP) optical+infrared, a \thco contour (blue) and \form contours (red) as in Figure~\ref{fig:http_highres}.  
    Spitzer-identified MYSOs are indicated with cyan circles, and the yellow contour indicates 230GHz continuum at 10$^{-4}\;$Jy$\;$bm$^{-1}\;\simeq\;$3$\sigma$. 
    The beam is shown in red in the lower left. The right panels show $T_K$ calculated from \form ratio 0.5*(\ltwo + \lthr)/\lone using the "simple" prescription at fixed N(\form) and n(H$_2$).  Only pixels with signal-to-noise $>$3 have been included.  There is modest evidence for higher line ratio consistent with external heating in the south-eastern pillars and central clumps, but instead the eastern clump "E-MYSO" shows evidence of internal heating.  Overlaid on the $T_K$ maps are contours of \lone at 30, 60, 110 and 170 mJy$\;$km$\;$s$^{-1}$, approximately 3,12,20 and 30$\sigma$ in the integrated line image.
    }
    \label{fig:http_ratio_zooms}
\end{figure}

\clearpage
\label{sec:zoom}
The higher resolution observations of 30Dor-10 offer the possibility to search for temperature structure within the 1$\farcs$75 clumps. Figure~\ref{fig:http_highres} shows the region observed at high resolution, with the distribution of \thco and \form emission relative to optical-infrared sources. 
At the full 0$\farcs$4 resolution, the signal-to-noise is not high enough to clearly measure temperature gradients within clumps.  Many clumps show evidence of a higher \ltwo/\lone and \lthr/\lone ratio on the edges of clumps than in the center, but that could also be a result of low signal-to-noise \lone emission in the denominator.  The trend is somewhat clearer after convolving the images to 0$\farcs$6 resolution: Fig.~\ref{fig:http_ratio_zooms} zooms into three regions, and along with the optical, \twco, and \form composite, temperature derived from the \form line ratio is shown.  
Integrated intensity maps were created from the cubes including all pixels in the cube greater than 2$\sigma$=2$\times$5$\;$mJy$\;$bm$^{-1}$.  The resulting noise in the integrated intensity images is 5$\;$mJy$\;$bm$^{-1}\;$km$\;$s$^{-1}$ for \lone and $\sqrt 2$ less in the averaged \ltwo and \lthr.  The ratio of the averaged higher lines to the lower lines was calculated for all pixels with $>$3$\sigma$ in both integrated intensity images.  Temperature was calculated from that ratio using the "simple" prescription of \citet{tang21} at n(H$_2$)=2$\times$10$^5\;$cm$^{-3}$ and N(\form)=10$^{13}\;$cm$^{-2}$.  We explored different options of masking thresholds, and although the magnitude of the temperature gradients changes moderately (high values change from $\gtrsim$180K to $\sim$160K, low values do not change significantly), the morphology is robust, as are the indications of internal and external heating discussed in the next paragraphs.

The first row \red{of Figure~\ref{fig:http_ratio_zooms}} shows the south-eastern pillars P1 and P2; these clumps are only 15~pc in projected distance from R136, and the \red{gas} density is very low south of this 30Dor-10 cloud\footnote{Most gas tracing emission (centimeter continuum, H$\alpha$, CO, etc) is very faint between R136 and the region studied here, and detailed modeling of optical emission lines shows that the region has negligible optical depth and an electron density of $\lesssim$50$\;$cm$^{-3}$ \citep{pellegrini}.  This cavity is bright in X-ray emission and modeled to be filled with N(H)$\sim$5$\times$10$^{21}\;$cm$^{-2}$ hot plasma \citep{townsley}.}, so there is very little attenuation of the ionizing radiation.
The \form clumps are surrounded by diffuse emission which is also very bright 
in H$\alpha$ and cm emission and thus most likely high emission measure ionized gas surrounding the molecular clumps. 
\citep[That emission has very similar distribution to the bright white emission in Figure~\ref{fig:http_highres}. See also ][Fig.~3c.]{indebetouw20}
The higher \form ratio on the edges of these clumps is consistent with exterior heating either from R136's radiation or by the surrounding ionized gas. 

The center region shown in the middle row contains two massive clumps, the southern of which has an embedded infrared source.  However, just to the west of those clumps lies a partially exposed small young cluster, as discussed in \citet{indebetouw20}.  
These clumps also show some evidence for higher line ratio on the outside, and possible exterior heating.  Interestingly, the clump containing the infrared source, in the center of the zoomed field, does not show clear evidence of internal heating.  Perhaps that source is too young or not high enough luminosity to have affected the bulk of the \form-emitting gas in this clump.

By contrast, the eastern clump "E-MYSO" in the bottom row shows no strong evidence for external heating, and instead some evidence of a higher temperature in the interior.  The lack of external heating is not unexpected, as that clump is located further back, evidently fairly shielded from direct radiation from R136.  One can assess the possibility that the embedded Massive Young Stellar Object (MYSO) is heating the cloud:  the \form-derived temperature is 35-40K on the cloud edges, and 75K in the center.  Interestingly, the excitation temperature derived from \twelveco \citep{wong22} is also 35-40K, albeit at the lower 1$\farcs$75 resolution.  This is consistent with an interpretation that the optically thick \twelveco traces cooler outer parts of the clump and \form traces warmer inner parts.  The H$_2$ column density derived from $^{12}$CO and $^{13}$CO$\;$2-1, assuming an abundance of N($^{13}$CO)/N(H$_2$) of 5$\times$10$^6$, is 3.6$\times$10$^{23}\;$cm$^{-2}$ \citep{wong22}.  If that column is distributed along a path length less than or similar to 0.5$\;$pc the size of the clump, the mean volume density is at least 2$\times$10$^5\;$cm$^{-3}$, and one can assume that the gas and dust temperatures are fairly well collisionally coupled. 

We fit the spectral energy distribution of this MYSO with the \citet{rob07} fitter \citep{zenfitter} and spubsmi grid of MYSO dust radiative transfer models \citet[][\url{https://doi.org/10.5281/zenodo.166732}]{rob17} to calculate a source luminosity of log(L)=4.7$\pm$0.3$\;$log(L$_\odot$).
We retrieved the output of the best-fitting models from \url{https://doi.org/10.5281/zenodo.572233} and find that the average dust temperature over the inner 0.1pc radius of these models is $\sim$80K.  Given the relatively high estimated densities, one would expect similar gas temperatures, agreeing with the temperatures observed in \form, especially considering that the \citet{rob17} models have a particular assumed density distribution.

\begin{figure}
\centerline{\resizebox{5in}{!}{\includegraphics{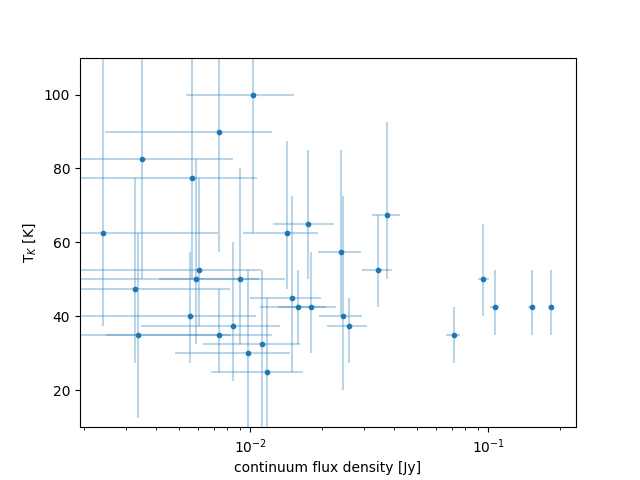}}}
\caption{$T_K$ fitted from three \form transitions, as a function of 232GHz continuum flux density.  The continuum is a combination of dust and free-free emission, but if \form heating were dominated by embedded star formation in clumps, one would expect some correlation, which is not evident.}
    \label{fig:continuum}
\end{figure}

As a further test for possible internal heating of clumps, we compare the \form-derived temperature $T_K$ to 232GHz continuum emission.  That continuum is a combination of dust and free-free, which cannot be distinguished without observations at other frequencies.  Nevertheless, star formation embedded in a clump can reasonably be expected to either heat dust, ionized gas, or both, and correlate with $T_K$, if that internal feedback dominates.  Figure~\ref{fig:continuum} shows no such correlation, supporting the hypothesis that external heating dominates in this region. 

\clearpage

\section{\form, CO, and feedback mechanisms}
\label{discussion}

\begin{figure}
    \centering
    \resizebox{3in}{!}{\includegraphics{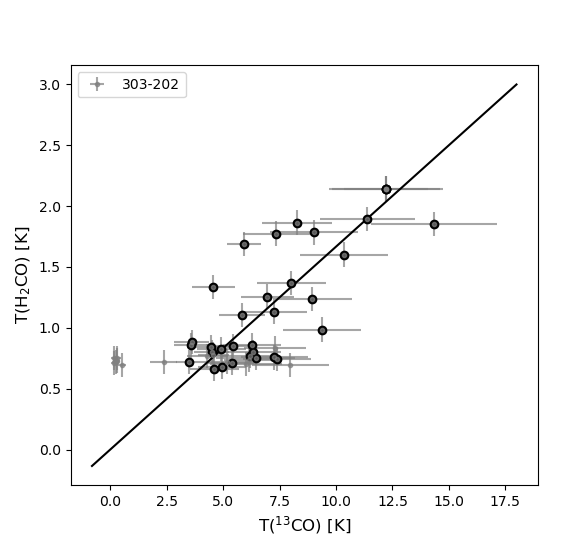}}
    \resizebox{3in}{!}{\includegraphics{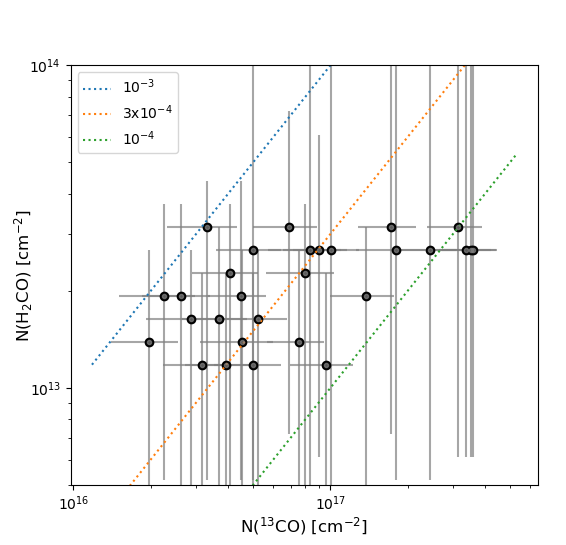}}
    \caption{\form \lone and \thco2-1 emission in clumps.  The first panel shows peak brightness temperature, showing some correlation.  The line is not a fit, but drawn at T(\form \lone) = 1/6 T(\thco2-1).  The second panel shows the fitted column densities; the uncertainties of N(\form) are too large to definitively detect any correlation, but the \thco/\form abundance ranges from 10$^{-4}$ to 10$^{-3}$ as indicated. See text for details.}
    \label{fig:h2co_13co}
\end{figure}

It is evident that \form is only detected in regions of the brightest CO emission. As shown in Figure~\ref{fig:h2co_13co}, the peak brightness temperatures of \form \lone and \thco$\;$2-1 are correlated, with T(\form\lone) $\simeq$1/6 T(\thco2-1).  If both lines are optically thin and have the same excitation temperature this would imply a correlation between the species' column densities $N$.  Unfortunately N(\form) is not very well constrained by non-LTE modeling of these lines.  The second panel \red{of Fig.~\ref{fig:h2co_13co}} shows those fitted column densities and the \thco column densities determined in \citet{wong22}, under the assumption that \twco and \thco are both in Boltzmann statistical equilibrium at the same excitation temperature, and that \twco is so optically thick that its brightness temperature reveals that excitation temperature without needing to know the \twco optical depth.  The uncertainties on N(\form) are too large to definitively say whether the two column densities are correlated, but the \form/\thco abundance ratio ranges between 10$^{-4}$ and 10$^{-3}$.  There are no clear trends of that abundance ratio in the spatial distribution or projected distance to R136, and the calculated column densities agree very well with those found by \citet{tang21} in the LMC star forming regions N159 and N113 (1-4$\times$10$^{13}$cm$^{-2}$ depending on the adopted volume density).
An assumed abundance ratio of and $H_2$/\thco=10$^6$ \citep[][and refs. therein]{indebetouw20} places our abundances in the typical range measured in cold clouds of 10$^{-10}$ to 10$^{-9}$ \citep{johnstone03,ao13}.  One possibility for elevated kinetic temperatures in \dor is X-ray heating, which can also affect the \form abundances.  Some authors have found 1-2 orders of magnitude enhancements of \form in X-ray intense regions like the Milky Way center \citep[e.g.][]{harada15}, but those regions also have high cosmic ray and mechanical heating, making it difficult to implicate one heating mechanism over another, and in contrast \citet{ao13} do not find significant abundance increases accompanying increased kinetic temperatures in the Galactic Center. 
\citet{liu20} present chemical models exploring the effect of X-rays on molecular clouds and find only modest increases of \form abundances: \form can form in the gas phase as well as on grain surfaces; although high X-ray irradiation can increase grain surface formation of \form by increasing the hydrogenation rate, the X-rays can also depress the gas phase abundances by increasing destructive reactions with more abundant cations. 
An assumed ratio of \twco/\thco=50 places our \form/\twco $\sim$ 10$^{-5}$ at the low end of the models in \citet{liu20}. 


\begin{figure}
    \centering
    \resizebox{3.5in}{!}{\includegraphics{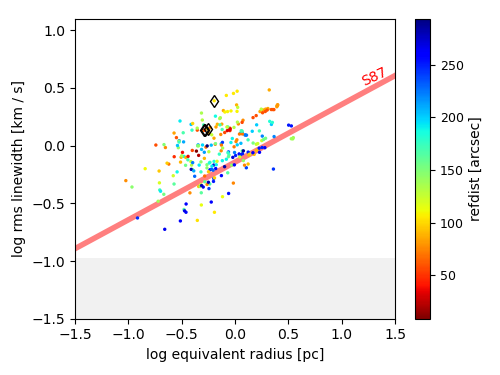}}
    \resizebox{3.5in}{!}{\includegraphics{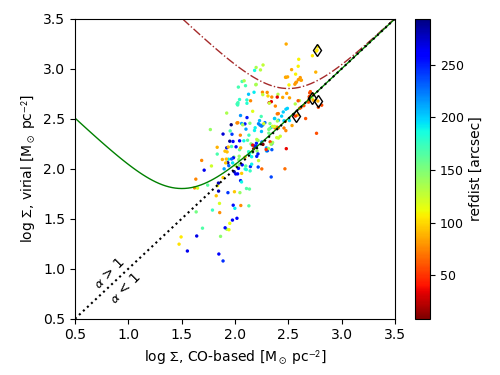}}
    \caption{Physical properties of \thco2-1 structures in 30Doradus, calculated identically to \citet{wong22}.  The first panel shows the velocity linewidth as a function of size, colored by projected distance to R136, and overlaid with the \citet{solomon87} size-linewidth relation.  The second panel compares the virial and LTE-determined surface densities.  On both, the \form-emitting clumps are highlighed with diamonds.  \form is detected in the higher column density clumps closer to R136.}
    \label{fig:size_linewidth}
\end{figure}

The size-linewidth-surface density properties of CO-emitting clumps in \dor were studied extensively in \citet{wong22}.  Figure~\ref{fig:size_linewidth} reproduces Figures of \thco clumps from that paper, but now with \form emitting clumps marked.  As noted above \form is associated with the higher column density clumps, and tend to be fairly close to R136, since the \thco column densities reach higher values in that central region.  The \form-emitting clumps are largely those with higher velocity linewidth $\sigma_v$, and lower gravitational ratios $\alpha=\Sigma_{vir}/\Sigma$.  

\citet{ao13} and \citet{ginsburg16} analyzed \form in the Milky Way central molecular zone and although they do not find a clear systematic trend with distance from the galactic center, they do find that all the warmest clouds are near the center.  They discuss four possible heating mechanisms: photoelectric, X-ray, cosmic ray, and turbulent/mechanical, and provide analytic estimates of each.  They find that turbulent or mechanical heating can most easily reach the 50-100K temperatures they observe (somewhat higher than in these LMC observations).  In particular \citet{ginsburg16} show that a strong correlation between gas temperature and nonthermal linewidth is expected in regions dominated by mechanical heating.

\begin{figure}
    \centering
    \resizebox{3in}{!}{\includegraphics{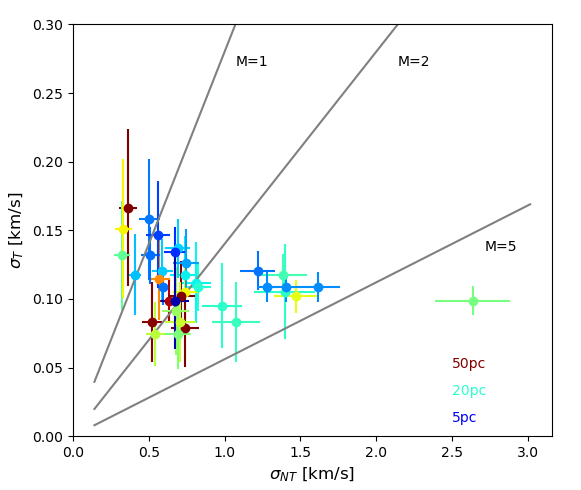}}
    \resizebox{3in}{!}{\includegraphics{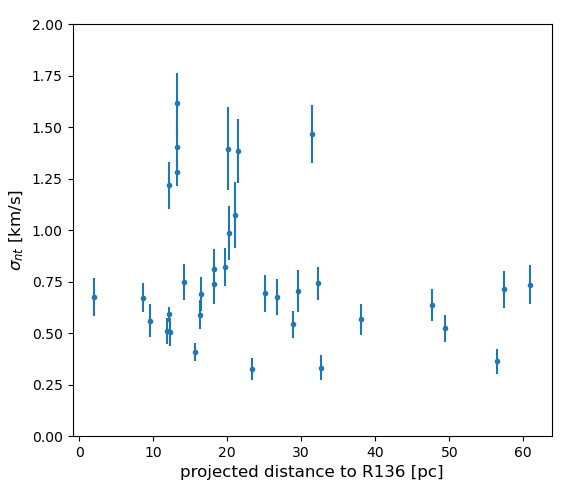}}
    \caption{(a) Thermal and nonthermal velocity dispersion of \form~\lone-emitting clumps.  Lines of~ constant Mach number 1,2, and 5 are labeled.  The clumps are colored according to projected distance from R136 on a rainbow color scale with several distances indicated. There is no apparent correlation between thermal and nonthermal energy in these clumps.  (b) Nonthermal velocity as a function of projected distance from R136. No strong trend is seen in these data.}
    \label{fig:h2co_mach}
\end{figure}

In LMC regions N113 and N159, \citet{tang21} found such a correlation between the fitted kinetic temperature and the nonthermal linewidth, suggesting that in those regions, the \form-emitting gas is being heated mechanically. Those two regions have a less significant main-sequence massive stellar population than 30Dor, and we have already found a correlation in 30Dor of temperature and proximity to those massive stars (primarily R136). Figure~\ref{fig:h2co_mach} shows the thermal and nonthermal velocity dispersions of the \form-emitting clumps, 
\begin{eqnarray*}
\sigma_T &=& \sqrt{kT_K/m(H_2CO)}\\
\sigma_{NT} &=& \sqrt{\sigma_v^2(3_{03}-2_{02})-\sigma_T^2}.
\end{eqnarray*}
The velocity widths of most clumps are dominated by the nonthermal component i.e. the Mach number is greater than 1,
\begin{equation*}
M=\sigma_{NT}/c_s = \sigma_{NT}/\sigma{T} \sqrt{30/2.37} >1.
\end{equation*}  
There is no clear correlation between the thermal and nonthermal linewidths.  Perhaps \dor is more dominated by external heating from the massive stellar radiation (photoelectric or X-ray heating), compared to N113 and N159, in which nonthermal motions may be providing a greater contribution to the gas heating.
Figure~\ref{fig:h2co_mach} also shows the nonthermal component of the linewidth as a function of projected distance from R136. Mechanical feedback, if dominated by R136, might also show a decrease with distance from the cluster, but no significant trend is present in these data. 

The overall X-ray luminosity of \dor of about 5$\times$10$^{36}$erg$\;$s$^{-1}$ \citep{townsley06} is quite similar to that in the Milky Way central molecular zone, leading to similar conclusions as \citet{ao13} that X-ray heating probably does not dominate, and would only raise the temperature to $\lesssim$10K.  \citet{townsley06b} isolated the  luminosity of R136 in particular, to be 2$\times$10$^{35}$erg$\;$s$^{-1}$, and the direct X-ray flux from R136 even on relatively unshielded gas at 10pc distance is even lower than the regional average. 

We can obtain a crude estimate of the cosmic ray heating in \dor as follows:  \citet{abdo10} fit Fermi data with several sources including \dor, to which they attribute a $\gamma$-ray energy flux F$_\gamma$ of 4$\times 10.^{-11}\;$erg$\;s^{-1}\;$cm$^{-2}$, but they can only localize that to a 4'=60pc radius region. \citet{schuppan} describe in detail how under the hadronic assumption, that $\gamma$-ray emission is primarily due to $\pi^0$ decay from p-p collisions, the $\gamma$-ray spectrum can be modeled to fit the energetic proton distribution, and then derive the ionization rate from those protons.  The factor is approximately 1.6$\times$10$^{54}$ times the $\gamma$-ray flux in erg$\;$s$^{-1}\;$cm$^{-2}$ times the distance to the region in kpc, squared, divided by the emitting volume in cm$^3$.  The scaling varies by a factor of a few depending on the spectral shape; \citet{abdo10} fit the \dor spectrum with a spectral index of 2 and upper energy cutoff of 7GeV, similar to the supernova remnants modeled by \citet{schuppan}.  An even larger source of uncertainty is that the conversion of $\gamma$-ray flux to ionization rate is uncertain by an order of magnitude because the Fermi-observable $\gamma$-rays originate from protons with $\gtrsim$1GeV protons, whereas gas ionization is dominated by much lower energy particles, so a significant assumption must be made to extrapolate the proton energy distribution to lower energies.
Nevertheless, this method derives an estimated $\zeta_I\sim$2$\times$10$^{-15}\;$s$^{-1}$ for the r=60pc \dor region.
Using \citet{ao13} Eqn.15 suggests this cosmic ray flux could heat the gas to $\sim$35K, so plausible within the large uncertainties.  The radial gradient centered near R136 would require the star cluster to be a localized source of cosmic rays.  So far the modest resolution of $\gamma$-ray observations cannot support or deny this for R136, but it has long been recognized that massive stellar wind termination shocks and wind-wind collisions can accelerate energetic particles \citep[][and refs therein]{benaglia,wang22}.

Photoelectric heating is the remaining viable heating mechanism, and it is generally dominant only in the outer photodissociated parts of a molecular cloud.  However, the \dor environment is unusual compared to molecular clouds in more quiescent environments.  The CO-dark molecular gas fraction in this central region close to R136 is likely $>$90\% \citep{chevance_dor}, and the clouds have complex geometries with high porosity and reduced dust abundance compared to solar-metallicity clouds.  Thus, we suggest that it is reasonable that in this extreme environment, the photodissociated, radiatively heated fraction of the cloud is quite large, and that elevated temperatures in the brightest \form-emitting gas could be due to radiative feedback from R136.

\section{Summary}

ALMA observations of \form emission at 0.4pc resolution in \dor were analysed to measure the gas kinetic temperature.  A trend of temperature decreasing from $\sim$60K down to $\sim$20K is found over the projected distance range of 10 to 30pc from R136.  In a subset of clouds observed at 0.1pc resolution, no strong signs of internal temperature gradients were found, despite some of those clouds containing embedded massive young stellar objects.  Of that subset, the clouds closer to R136 do show evidence of higher temperature on the outsides, consistent with external heating.

Assessing various heating mechanisms, we do not observe any correlation between the clump linewidth and the kinetic temperature or distance from R136, as might be expected for mechanical heating dominated by turbulence.  Rough estimates of the magnitude of X-ray heating fall short of the observed temperatures traced by \form, as has been found in other regions.  Within very large uncertainties about the assumed proton spectrum, cosmic ray heating could achieve the observed gas temperatures; more precise determinations of the cosmic ray density, perhaps derived from astrochemistry, could provide a more definitive conclusion.  The radial temperature gradient would require that the massive stellar winds in R136 (wind-wind and wind termination shocks) are the dominant source of cosmic rays in the region, which would constitute an interesting result.
We suggest that the \form-emitting gas in these clouds near R136 may also be dominated by the photoelectically-heated PDR gas phase, and the observed gas temperature might be due primarily to the direct radiative feedback from R136.  Higher resolution PDR modeling than was possible with Herschel will help elucidate the physics of this feedback-dominated region.

\section{Acknowledgements}

This research made use of astrodendro, a Python package to compute dendrograms of Astronomical data (http://www.dendrograms.org/).  
This paper makes use of the following ALMA data: ADS/JAO.ALMA\#2019.1.00843.S and ADS/JAO/ALMA\#2013.1.00346.S.
ALMA is a partnership of ESO (representing its member
states), NSF (USA) and NINS (Japan), together with NRC
(Canada), NSC and ASIAA (Taiwan), and KASI (Republic of
Korea), in cooperation with the Republic of Chile. The Joint
ALMA Observatory is operated by ESO, AUI/NRAO, and
NAOJ. The National Radio Astronomy Observatory is a
facility of the National Science Foundation operated under
cooperative agreement by Associated Universities, Inc.
T.W. and R.I. acknowledge support from collaborative NSF AAG awards 2009849 and 2009624.
M.S. acknowledges partial support from the NASA ADAP Grant Number 80NSSC22K0168. The material is based upon work supported by NASA under award number 80GSFC21M0002 (M.S.).
M.C. gratefully acknowledges funding from the DFG through an Emmy Noether Research Group (grant number CH2137/1-1). COOL Research DAO is a Decentralized Autonomous Organization supporting research in astrophysics aimed at uncovering our cosmic origins.
\red{M.R. wishes to acknowledge support from ANID(CHILE) through FONDECYT grant No1190684 and partial support from ANID Basal FB210003.}

\bibliographystyle{aasjournal}
\bibliography{h2co30dor.bib}

\end{document}